\documentclass[12pt]{article}
\usepackage{amsmath,amssymb,color}
\usepackage{cite,epsf}
\usepackage{pstricks}
\usepackage{color}
\usepackage{hyperref}
\usepackage{tikz}     

\usetikzlibrary{
decorations.pathmorphing 
}

\usepackage{graphicx,array} 
\newcommand{\be}{\begin{equation}}
\newcommand{\ee}{\end{equation}}
\newcommand{\beq}{\begin{equation}}
\newcommand{\eeq}{\end{equation}}
\newcommand{\bea}{\begin{eqnarray}}
\newcommand{\eea}{\end{eqnarray}}

\newcommand{\ba}{\begin{eqnarray}}
\newcommand{\ea}{\end{eqnarray}}

\def\sin{\mbox{sin}}
\def\cos{\mbox{cos}}

\def\log{\mbox{log}}

\begin{document}

\begin{titlepage}
\vspace{10pt}
\hfill
{\large\bf HU-EP-18/33}
\vspace{20mm}
\begin{center}

{\Large\bf  On Wilson loops for two touching circles \\with opposite orientation

}

\vspace{45pt}

{\large Harald Dorn 
{\footnote{dorn@physik.hu-berlin.de
 }}}
\\[15mm]
{\it\ Institut f\"ur Physik und IRIS Adlershof, 
Humboldt-Universit\"at zu Berlin,}\\
{\it Zum Gro{\ss}en Windkanal 6, D-12489 Berlin, Germany}\\[4mm]

\vspace{20pt}

\end{center}
\vspace{10pt}
\vspace{40pt}

\centerline{{\bf{Abstract}}}
\vspace*{5mm}
\noindent
We study the Wilson loops for contours formed by a consecutive passage of two touching circles with a common tangent, but opposite orientation. The calculations are 
performed in lowest nontrivial order for ${\cal N}=4$ SYM at weak and strong coupling and for QCD at weak coupling.
After subtracting the
standard linear divergence proportional to the length, as well the recently
analysed spike divergence, we get for the renormalised Wilson loops $\mbox{log}~W_{\mbox{\scriptsize ren}}=0$. The result holds for circles with different radii and arbitrary angle between the discs spanned
by them.

\vspace*{4mm}
\noindent

\vspace*{5mm}
\noindent
   
\end{titlepage}
\newpage


\section{Introduction}
 Ultraviolet divergences of Wilson loops for smooth contours, as well as for those with cusps and intersecting points,  have been studied in much detail
from the early eighties to present time. Especially the cusp anomalous dimension has drawn a lot of attention since it  is also related to various other
physical situations, see e.g. \cite{Grozin:2015kna} and references therein. It diverges in the limit of vanishing opening angle. However, the removal of the regularisation does not commute with
that limit, and only recently we have started the investigation of renormalisation in the presence of zero opening angle cusps. i.e. spikes \cite{Dorn:2018als}.

A spike turned out to be responsible for a divergence proportional to the inverse of the square root out of the product of the dimensional cutoff times
the jump in the curvature. The analysis has been performed in lowest order at weak coupling both for ${\cal N}=4$ SYM and QCD and at strong
coupling via holography in the supersymmetric case. In addition, the spike generates in the SUSY case, at least at weak coupling, an additional logarithmic
divergence, which could be related to the breaking of   zig-zag symmetry \cite{Polyakov:1997tj}, \cite{Drukker:1999zq}. 

Although the lowest order setting in \cite{Dorn:2018als} was very simple, the safe extraction of terms beyond the leading divergence required some technical 
effort. In the present paper we go one step further and want to evaluate also the finite terms, which after subtraction of the divergences define
the renormalised Wilson loops. This we will do for a special contour. It is formed out of two touching circles with a common tangent in the following way.
After starting at the common point one traverses the first circle and then continues along the second circle in just the opposite direction. The discs related to the circles are allowed to form an angle $\beta$.

The paper is organised as  follows. The next section is devoted to lowest order at weak coupling for the locally  supersymmetric Wilson loop. Then
section 3 contains the holographic analysis at strong 't Hooft coupling. In section 4 we comment on the situation without supersymmetry
by subtracting the scalar contributions from the result in section 2. After the concluding section follow two appendices containing
the technical details of the asymptotic estimates of the necessary integrals.

\section{Lowest order at weak coupling in ${\cal N}=4$ SYM}
In ${\cal N}=4$ SYM the Euclidean local supersymmetric Wilson loop for a closed contour parameterised by $x^{\mu}(\tau)$ is given by
\cite{Rey:1998ik, Maldacena:1998im}, \cite{Drukker:1999zq}
\beq
W~=~\frac{1}{N}\Big \langle \mbox{tr Pexp}\int \big (iA_{\mu}\dot x^{\mu}~+~\vert \dot x\vert \phi_I\theta^I\big )~d\tau\Big \rangle~.\label{Wloop}
\eeq
For simplicity we consider only the case of fixed $\theta ^I\in S^5$. 

Our contour of interest  has been characterised in the introduction. Let the two circles 
with radii
\beq 
R_1~>~R_2\label{R1geqR2}
\eeq
be parameterised by
\bea
\vec x_1(\varphi_1)&=&R_1\big(\sin\varphi_1,~1-\cos\varphi_1,~0\big)~,\nonumber\\
\vec x_2(\varphi_2)&=&R_2\big (\sin\varphi_2,~\cos\beta(1-\cos\varphi_2),~\sin\beta(1-\cos\varphi_2)\big )~.\label{x-circles}
\eea
Then the contour to be  used in \eqref{Wloop}
is given by
\bea
\vec x(\tau)&=&\vec x_1(\tau)~,~~~~~~~~~~~~0\leq\tau\leq2\pi~,\nonumber\\
\vec x(\tau)&=&\vec x_2(4\pi-\tau)~,~~~~2\pi\leq\tau\leq 4\pi~.\label{contour}
\eea
The situation for a fixed larger circle and smaller partner circles at various values of the angle $\beta$ is illustrated in figure 1.\\
\begin{figure}[h!]
 \centering
\includegraphics[width=10cm]{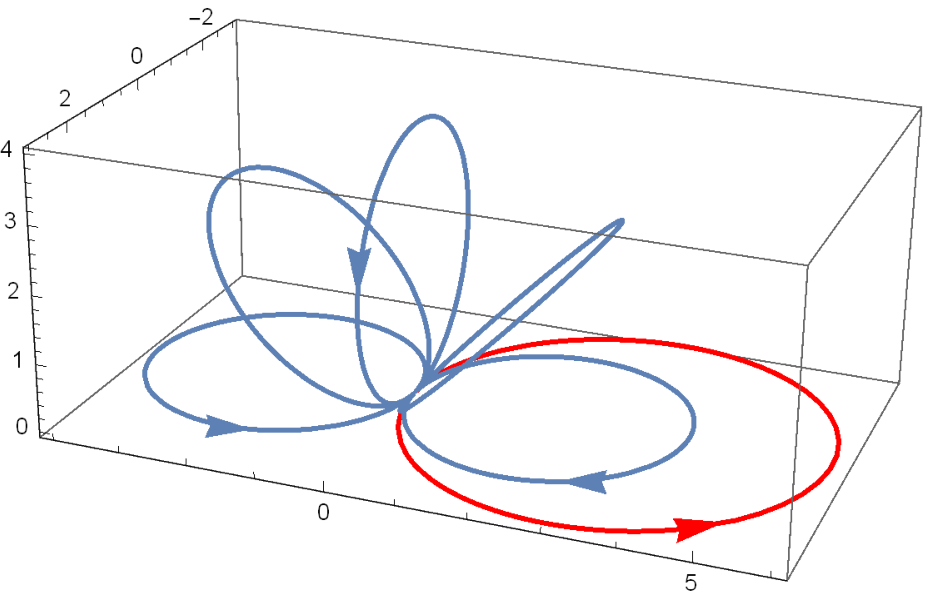}
\\[5mm]
Figure 1: {\it Larger circle (red)  with radius $R_1=3$  and smaller \\$~~~~~~~~~~~~~~~~$ circle (blue) with $R_2=2$ at various angles $\beta\in [0,\pi]$.}  
\label{fig1}
\end{figure}

Then the perturbative expansion of this Wilson loop is given by
\beq
\mbox{log}W~=~\frac{g^2C_F}{4\pi^2}\Big (I_{1}~+~I_{2}~+~I_{12}\Big )~+~{\cal O}(g^4)~.
\label{logW}
\eeq
The integrals $I_1$ and $I_2$ correspond to the contributions, where both endpoints of the propagators  are on the same circle. In $I_{12}$ 
 the propagators connect the two circles. This means ($\epsilon$ denotes a dimensionful parameter for UV regularisation)
\beq
I_{j}=\int_0^{2\pi}\int_0^{\varphi_1}\frac{R_j^2(1-\cos(\varphi_1-\varphi_2))~d\varphi_1d\varphi_2}{2R_j^2(1-\cos(\varphi_1-\varphi_2))+\epsilon^2}~.\label{Ij}
\eeq
Performing one trivial integration one gets
\beq
I_{j}~=~\frac{\pi}{2}\int_0^{2\pi}\frac{(1-\cos x)~dx}{1-\cos x+\frac{\epsilon^2}{2R_j^2}}~=~\pi^2~+~{\cal O}(\epsilon)~.\label{Ijfinal}
\eeq
Furthermore $I_{12}$ is given by
\beq
I_{12}=\int_0^{2\pi}\int_0^{2\pi}~\frac{R_1R_2(1+\cos\varphi_1\cos\varphi_2+\cos\beta~\sin\varphi_1\sin\varphi_2)~d\varphi_1~d\varphi_2}{
(\vec x_1~-~\vec x_2)^2~+~\epsilon^2}\label{I12original}~.
\eeq

Then, performing the $\varphi_2$-integration, we get \footnote{Since the remaining integrand depends on $\cos\varphi$ only, the original integral over $(0,2\pi)$ can be written as twice that over $(0,\pi)$. }
\beq
I_{12}~=~4\pi~\int_0^{\pi}~\big (f(R_1,R_2,\beta,\varphi)~+~g(R_1,R_2,\beta,\epsilon,\varphi)\big )~d\varphi
~,\label{I12fg}
\eeq
with
\beq
f(R_1,R_2,\beta,\varphi)~=~\frac{1}{2}~\frac{R_1(R_1\cos\beta(\cos\varphi-1)-R_2\cos\varphi)}{R_1^2\sin^2\varphi+(R_2+R_1\cos\beta(\cos\varphi-1))^2}\label{deff}\\[2mm]
\eeq
and, using the abbreviation
\bea
U(R_1,R_2,\beta,\varphi)&=&R_1R_2~\cos\beta(1-\cos\varphi)-R_2^2~,\nonumber\\[2mm]
g(R_1,R_2,\beta,\epsilon,\varphi)&=&\frac{R_1R_2-\frac{\big (\epsilon^2+2R_1^2(1-\mbox{\scriptsize cos}\varphi) -2 U\big )\big (R_1R_2U\mbox{\scriptsize cos}\varphi -R_1^2R_2^2\mbox{\scriptsize cos}\beta~ \mbox{\scriptsize sin}^2\varphi \big )}{2U^2+2R_1^2R_2^2 \mbox{\scriptsize sin}^2\varphi}}{\sqrt{(\epsilon^2+2R_1^2(1-\cos\varphi)-2U)^2-4U^2-4R_1^2R_2^2\sin^2\varphi}}~.\label{defg}
\eea
The indefinite integral over $f(R_1,R_2,\beta,\varphi)$ is
$$-~\frac{1}{2}~\mbox{arctan}\Big (\frac{R_1 \sin\varphi}{R_2-R_1\cos\beta(1-\cos\varphi)}\Big )~.$$
It is zero at both ends of the integration interval of the definite integral needed in \eqref{I12fg}.  However one has to be careful, since for
$$ \cos\beta~>~\frac{R_2}{2R_1}$$
the argument of $\arctan$-function passes infinity within the integration interval\\ $\varphi\in [0,\pi]$. This leads to ($\Theta$ denoting the step function)
\beq
\int_0^{\pi}f(R_1,R_2,\beta,\varphi )~d\varphi~=~-\frac{\pi}{2}~\Theta\Big (\cos\beta- \frac{R_2}{2R_1}\Big)~.\label{fint}
\eeq 

For the integral over $g(R_1,R_2,\beta,\epsilon,\varphi)$ we change the integration variable via
\beq
\frac{1}{2}~B_{\epsilon}^2(1-\cos\varphi)~=~x^2~,\label{phi-x}
\eeq
where we introduced  the abbreviations
\beq
B_{\epsilon}~=~\sqrt{\frac{2R_1R_{12}}{\epsilon~ R_2}}~\label{A}
\eeq
and
\beq
R_{12}(R_1,R_2,\beta)~=~\sqrt{R_1^2+R_2^2-2R_1R_2~\cos\beta}~.\label{R12}
\eeq
$R_{12}$ is just the distance between the centers of the two circles. It is also
via
\beq
\frac{R_{12}}{R_1R_2}~=~\vert \vec k_1-\vec k_2\vert \label{R12-curv}
\eeq
related to the difference of the curvature vectors at the touching point.

Then we arrive with \eqref{I12fg}, \eqref{defg}  and \eqref{fint} at
\beq
I_{12}~=~-2\pi^2~\Theta\Big (\cos\beta -\frac{R_2}{2R_1}\Big)~+~\frac{4\pi}{\sqrt{\epsilon}}~\sqrt{\frac{2R_1R_2}{R_{12}}}~
\int_0^{B_{\epsilon}}~\frac{h(R_1,R_2,\beta,\epsilon,x)~dx}{\sqrt{(1-\frac{x^2}{B_{\epsilon}^2})(1+x^4)}  }~.\label{I12x}
\eeq
In the above equation use has been made of the following definitions
\beq
h(R_1,R_2,\beta,\epsilon,x)=\frac{1+\frac{\epsilon^2}{4R_2^2}+h_1~\epsilon x^2+h_2~\epsilon^2x^4}{1+2h_3~\epsilon x^2-\frac{\mbox{\scriptsize sin}^2\beta}{R_{12}^2}\epsilon^2x^4}~\sqrt{\frac{1+x^4}{1+x^4+\frac{\epsilon^2}{4R_2^2}+h_3~\epsilon x^2}}~,\label{h}
\eeq
with
\bea
h_1(R_1,R_2,\beta,\epsilon)&=&\frac{R_2(4(R_1^2-R_2^2)+2R_{12}^2)+(R_1\mbox{ cos}\beta-R_2)\epsilon^2}{4R_1R_2^2R_{12}}~,\\[2mm]
h_2(R_1,R_2,\beta)&=&\frac{(R_1^2+R_2^2)~\mbox{cos}\beta-2R_1R_2}{2R_1R_2R_{12}^2}~,\\[2mm]
h_3(R_1,R_2,\beta)&=&\frac{R_1-R_2~\mbox{cos}\beta}{R_2R_{12}}~.\label{h3}
\eea

We are interested in the finite piece of $I_{12}$ at $\epsilon\rightarrow 0$. Therefore, we have to keep control
also over the ${\cal O}(\sqrt{\epsilon})$ contribution to the integral in \eqref{I12x}. Now for each fixed $x$ the nominator in the integrand of \eqref{I12x} is $h=1+{\cal O}(\epsilon)$. But, unfortunately, this estimate  does not hold uniformly in the
whole integration range $(0,B_{\epsilon})$. Hence the necessary analysis requires some detailed care and is put into appendix A. Inserting its result 
\eqref{appAfinal} for the integral into \eqref{I12x} we get \footnote{$o(\epsilon^0)$ denotes terms vanishing for $\epsilon\rightarrow 0.$}
\beq
I_{12}~=~\sqrt{\frac{2\pi R_1R_2}{R_{12}}}~\Big (\Gamma\big (\frac{ 1}{ 4}\big )\Big )^2~\frac{1}{\sqrt{\epsilon}}~-~2\pi^2~+~o(\epsilon^0)~.\label{I12final}
\eeq
As one should have expected, the discontinuities at $\cos\beta=\frac{R_2}{2R_1}$, i.e.$R_{12}=R_1$, showing up in both \eqref{fint} and \eqref{appAfinal}, cancel in the final result for $I_{12}$.

With \eqref{Ijfinal} and \eqref{I12final} into \eqref{logW} one gets, after subtraction of the $\frac{1}{\sqrt{\epsilon}}$ spike divergence\nolinebreak  \cite{Dorn:2018als},
\beq
\mbox{log}~W_{\mbox{\scriptsize ren}}~=~0~+~{\cal O}(g^4)~.
\eeq 
\section{Holographic evaluation at strong coupling}
To generate the two circles as the image of two straight lines after an inversion on the unit sphere, we have to choose for these lines
\bea
\vec y_1(\tau)&=&\Big (\tau,\frac{1}{2R_1},0\Big )~,\nonumber\\
\vec y_2(\tau)&=&\Big (\tau,\frac{\cos\beta}{2R_2},\frac{\sin\beta}{2R_2}\Big )~,\label{lines}
\eea
with
\beq
-\infty ~<~\tau ~<~\infty~.
\eeq
The distance between them is
\beq
L~=~\frac{R_{12}}{2R_1R_2}~,\label{L}
\eeq
with $R_{12}$ from \eqref{R12}.
\begin{figure}[h!]
 \centering
\includegraphics[width=10cm]{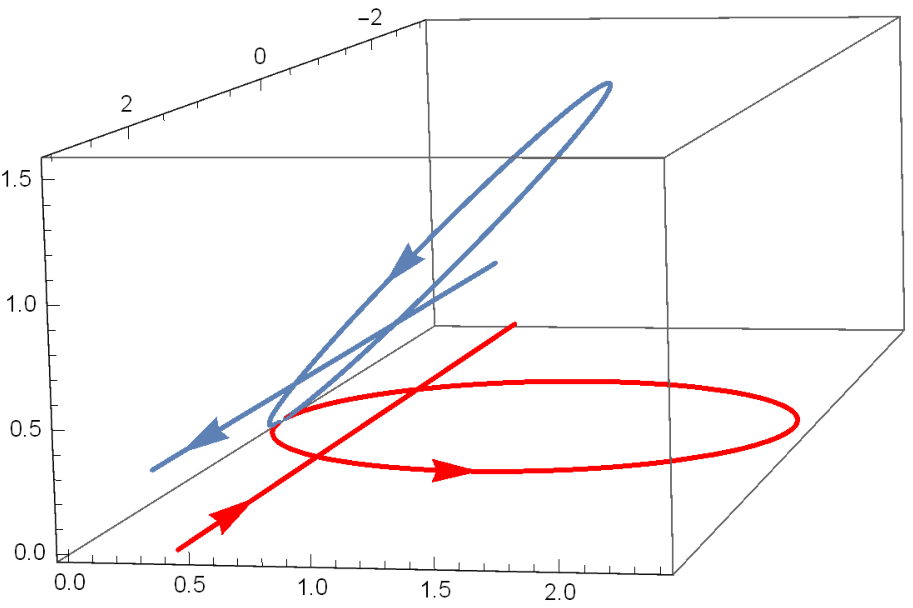}
\\[5mm]
Figure 2: {\it In red: Larger circle with radius $R_1=1.2$  and parts of its preimage. \\$~~~~~~~$In blue:  The same for smaller circle  with $R_2=1.1~,~\beta=\pi/4$.}  
\label{fig2}
\end{figure}

As a result one gets the circles in the form 
\bea
\vec x_1(\tau)&=&\frac{4R_1^2}{1+4R_1^2\tau^2}\Big (\tau,\frac{1}{2R_1},0\Big )~,\nonumber\\
\vec x_2(\tau)&=&\frac{4R_2^2}{1+4R_2^2\tau^2}\Big (\tau,\frac{\cos\beta}{2R_2},\frac{\sin\beta}{2R_2}\Big )~.\label{circles}
\eea

The minimal surface in $AdS$, approaching the two straight lines \eqref{lines} on the boundary, is given by (in Poincar$\acute{ \mbox{e}}$ coordinates $x_1,x_2,x_3,z,$ with $z=0 $ as boundary, $ds^2=(dz^2+dx_1^2+dx_2^2+dx_3^2)/z^2$) \cite{Maldacena:1998im}
\bea
z(\sigma,\tau)&=& r(\sigma)~,~~~~~~~~~~~~~~~~~~\sigma\in(-L/2,L/2)~,~~\tau\in(-\infty,\infty)~,\nonumber\\
x_1(\sigma,\tau)&=&\tau~,\nonumber\\
x_2(\sigma,\tau)&=&\frac{1}{2R_1}\Big (\frac{1}{2}-\frac{\sigma}{L}\Big )+\frac{1}{2R_2}\Big (\frac{1}{2}+\frac{\sigma}{L}\Big )~\cos\beta~,\nonumber\\
x_3(\sigma,\tau)&=&\frac{1}{2R_2}\Big (\frac{1}{2}+\frac{\sigma}{L}\Big )~\sin\beta~.\label{orig-surf}
\eea 
The function $r(\sigma)$ is defined via
\beq
r(-\sigma)~=~r(\sigma)~~~\mbox{and}~~~~\sigma~=~r_0~\int _{\frac{r(\sigma)}{r_0}}^1\frac{y^2dy}{\sqrt{1-y^4}}~~~\mbox{for}~0<\sigma<\frac{L}{2}~,\label{sigma-r}
\eeq
with $r_0$ fixed by
\beq
L~=~2r_0\int_0^1\frac{y^2dy}{\sqrt{1-y^4}}~=~\frac{(2\pi)^{3/2}~r_0}{\big (\Gamma(\frac{1}{4})\big)^2}~. \label{r0L}
\eeq
The AdS isometry
\beq
x_{\mu}~\mapsto~\frac{x_{\mu}}{x^2+z^2}~,~~~~~~z~\mapsto ~\frac{z}{x^2+z^2}~ \label{trafo}
\eeq
acts on the boundary $(z=0)$ as inversion on the unit sphere, mapping the straight
lines \eqref{lines} and circles \eqref{circles} to each another. Therefore, the minimal surface in $AdS$, approaching the two circles \eqref{circles} is given by the
image of \eqref{orig-surf} under the map \eqref{trafo}, i.e. by
\bea
\left( \begin{array}{c} x_1 \\ x_2 \\ x_3\\z \end{array} \right)&=&\Big (\tau^2+r^2(\sigma)+\frac{\sin^2\beta}{4R_{12}^2}+\big( \sigma +\frac{R_1^2-R_2^2}{4R_1R_2R_{12}}\big )^2\Big )^{-1}\nonumber\\
&&\cdot ~\left( \begin{array}{c} \tau\\\frac{1}{2R_1}\Big (\frac{1}{2}-\frac{\sigma}{L}\Big )+\frac{1}{2R_2}\Big (\frac{1}{2}+\frac{\sigma}{L}\Big )~\cos\beta  \\\frac{1}{2R_2}\Big (\frac{1}{2}+\frac{\sigma}{L}\Big )~\sin\beta   \\r(\sigma) \end{array} \right )~.\label{surf}
\eea
The regularised area $A_{\epsilon}$, needed for the holographic evaluation of our
Wilson loop, is then just the area of that part of \eqref{surf}, for which
$z>\epsilon$. Its boundary, as parameterised by $\sigma$ and $\tau$, is given by
\beq
\epsilon~=~\frac{r(\sigma)}{\tau^2+r^2(\sigma)+\frac{\mbox{\scriptsize sin}^2\beta}{4R_{12}^2}+\big( \sigma +\frac{R_1^2-R_2^2}{4R_1R_2R_{12}}\big )^2}~.\label{boundary}
\eeq
Based on the isometric character of the map \eqref{trafo}, we prefer as in \cite{Dorn:2018als} to calculate $A_{\epsilon}$ on the preimage \eqref{orig-surf}. There the induced metric is independent of $\tau$ and 
areas are given by $r_0^2\int\frac{d\tau d\sigma}{r^4(\sigma)}$. To change the integration variable from $\sigma$ to $r$ one has to keep in mind, that their relation is not one to one. Let $\sigma(r)\geq 0$ be given by the integral in \eqref{sigma-r}. Then we get
\beq
A_{\epsilon}~=~A_{\epsilon}^+~+~A_{\epsilon}^-~,\label{A-Apm}
\eeq
\beq
A_{\epsilon}^{\pm}~=~\int _{{\cal B}^{\pm}_{\epsilon}}~\frac{1}{r^2\sqrt{1-(\frac{r}{r_0})^4}}~dr d\tau~.
\eeq
The integration regions ${\cal B}^{\pm}_{\epsilon}$ are defined by
\beq
\frac{r}{\tau ^2+r^2+(M\pm \sigma(r))^2+\frac{\mbox{\scriptsize sin}^2\beta}{4R_{12}^2}}~>~\epsilon~,
\eeq
with \footnote{$M$ depends on $\beta$ via $R_{12}(R_1,R_2,\beta) $. For $\beta=0$ this agrees with the formulas in \cite{Dorn:2018als} of course.}
\beq
M(R_1,R_2,\beta)~=~\frac{R_1^2-R_2^2}{4R_1 R_2 R_{12}}~.\label{M-def}
\eeq 
Performing the trivial $\tau$-integration (see \cite{Dorn:2018als}) we arrive at
\beq
A_{\epsilon}^{\pm}~=~\frac{2~r_0^2}{\sqrt{\epsilon}}~\int_{r_{\epsilon}^{\pm}}^{r_0}\frac{\sqrt{r-\epsilon~ (M\pm \sigma(r))^2-\epsilon~ r^2-\epsilon~\frac{\mbox{\scriptsize sin}^2\beta}{4R_{12}^2} }}{r^2~\sqrt{r_0^4-r^4}}~dr~. 
\label{Aeps}
\eeq
The lower boundaries $r_{\epsilon}^{\pm}$ are defined as solutions of
\beq
r_{\epsilon}^{\pm}~-~\epsilon~\Big (\big (M\pm \sigma(r_{\epsilon}^{\pm})\big )^2+(r_{\epsilon}^{\pm})^2 +\frac{\sin^2\beta}{4R_{12}^2} \Big )~=~0~.\label{region}
\eeq
The evaluation of these integrals for $\epsilon\rightarrow 0$ up to divergent and ${\cal O}(\epsilon^0)$ terms is performed in appendix B. After applying some $\Gamma$-function arithmetic to the result \eqref{appBfinal}
we get 
\beq
A_{\epsilon}~=~\frac{2\pi(R_1+R_2)}{\epsilon}~-~ \frac{32\pi^{\frac{7}{4}}\sqrt{2
\sqrt{2}+3}}{(\Gamma(\frac{1}{8}))^2}~\frac{1}
{\sqrt{\epsilon\vert \vec k_1-\vec k_2\vert}}~+~{\cal O}(\sqrt{\epsilon})~.\label{Aeps-final}
\eeq
The leading divergent term is due to the standard $1/\epsilon$ divergence proportional 
to the length of the boundary contour. 
The next-leading $1/\sqrt{\epsilon}$ divergence is just twice the spike divergence
analysed in \cite{Dorn:2018als}. After subtracting these divergences the remainder
tends to zero for $\epsilon\rightarrow 0$, hence
\beq
A_{\mbox{\scriptsize ren}}~=~0~.
\eeq
Then via the holographic Wilson loop formula \cite{Maldacena:1998im} we get
at large $N$ and strong \\ 't Hooft coupling $\lambda=g^2 N$
\beq
\log W_{\mbox{\scriptsize ren}}~=~0~.
\eeq

Before closing this section we have to mention a certain subtlety. There is still another potentially competing surface, the disconnected
\footnote{Up to the touching point on the boundary of AdS.} one, built out of the surfaces for the two single circles.  First of all it is discriminated
by the fact, that the regularised contour generated by cutting at $z=\epsilon$ is not connected. Furthermore, its
regularised area is
\cite{Drukker:1999zq, Zarembo:1999bu}
\beq
A_{\epsilon}^{\mbox{\scriptsize disconn}}~=~\frac{2\pi(R_1+R_2)}{\epsilon}~-~4\pi~+~{\cal O}(\epsilon)~. \label{Adisconn}
\eeq
For applications to the holographic evaluation of Wilson loops the common leading $1/\epsilon$-divergence is cancelled by a boundary term induced by a necessary Legendre transformation \cite{Drukker:1999zq}. 
For small $\epsilon$ the disconnected surface is once more  discriminated, since \eqref{Adisconn} both as it stands as well as after subtraction of the leading term is larger than \eqref{Aeps-final}.  However it would win, if the values of the finite
pieces would have to decide. To my knowledge so far this alternative did not play any role in papers studying the Gross-Ooguri phase transition \cite{Gross:1998gk}, since there the competing  areas had the same divergent parts. Only in a recent paper \cite{Munkler:2018cvu} on  the cross anomalous dimensions a comparison of areas
with differing divergent terms was relevant and the decisions were based also on the full regularised areas.
\section{Comment on the ordinary  Wilson loop} 
The ordinary  (not supersymmetric)  Wilson loop is given by \eqref{Wloop} without the coupling of the contour to the scalars. According to the recipe for its
holographic evaluation, as formulated in \cite{Alday:2007he,Polchinski:2011im}, at leading order strong coupling it coincides with the
supersymmetric Wilson loop as studied in the previous section. 

To handle the leading order at weak coupling, we have to subtract the scalar contributions  
from those in section 2. The result is then valid both for the ordinary Wilson loop in ${\cal N}=4$ SYM and QCD.

There are the two trivial terms with both points of the propagator on the same circle 
\beq
I_j^{\mbox{\scriptsize scalar}}~=~\frac{\pi}{2}\int_0^{2\pi}\frac{dx}{1-\cos x+\frac{\epsilon^2}{2R_j^2}}~=~\frac{\pi^2 R_j}{\epsilon}~+~{\cal O}(\epsilon)~.\label{Ij-scalar}
\eeq
For
\beq
I_{12}^{\mbox{\scriptsize scalar}}~=~\int_0^{2\pi}\int_0^{2\pi}~\frac{R_1R_2~d\varphi_1~d\varphi_2}{
(\vec x_1~-~\vec x_2)^2~+~\epsilon^2}\label{I12scalar-orig}~
\eeq 
we get after performing the $\varphi_2$-integration \footnote{To keep formulas short, we write down only the case $\beta =0$. Then $R_{12}=R_1-R_2$.}
\beq
I_{12}^{\mbox{\scriptsize scalar}}=4\pi\int_0^{\pi}\frac{R_1R_2~d\varphi}{\sqrt{\epsilon^4+4\epsilon^2R_2^2+4\epsilon^2R_1R_{12}(1-\cos\varphi)+4R_1^2R_{12}^2(1-\cos\varphi)^2}}~.
\eeq
After the change of integration variable as indicated in \eqref{phi-x} this becomes
\beq
I_{12}^{\mbox{\scriptsize scalar}}~=~\frac{2\pi}{\sqrt{\epsilon}}~\sqrt{\frac{2R_1R_2}{ R_{12}}}\int_0^{B_{\epsilon}}\frac{dx}{\sqrt{\big (1-\frac{x^2}{B_{\epsilon}^2}\big )
\big (1+x^4+\frac{\epsilon^2}{4R_2^2}+\frac{\epsilon x^2}{R_2}\big )}}~.
\eeq
Now an analysis analogously to appendix A yields
\beq
I_{12}^{\mbox{\scriptsize scalar}}~=~\sqrt{\frac{\pi R_1R_2}{2R_{12}}}~\Big (\Gamma(\frac{1}{4})\Big )^2\frac{1}{\sqrt{\epsilon}}~+~o(\epsilon ^0)~.\label{I12-scalar}
\eeq
Note that both \eqref{Ij-scalar} and \eqref{I12-scalar} beyond the divergent terms
contain no finite term remaining in the limit $\epsilon\rightarrow 0$.

The QCD Wilson loop becomes
\beq
 \mbox{log}~W^{\mbox{\scriptsize QCD}}~=~\frac{g^2C_F}{4\pi^2}\left (-\frac{\pi^2(R_1+R_2)}{\epsilon}+\sqrt{\frac{\pi R_1R_2}{2R_{12}}}~\Big (\Gamma(\frac{1}{4})\Big )^2\frac{1}{\sqrt{\epsilon}}+o(\epsilon ^0)\right )+{\cal O}(g^4)~.
\eeq
Then after subtraction of the standard $1/\epsilon$ divergence proportional to the length and the QCD spike divergence \cite{Dorn:2018als} our final result for the renormalised Wilson loop is \footnote{Since for lowest order weak coupling the scalars contribute no finite nonvanishing term, log$W_{\mbox{\scriptsize ren}}$=0 holds in this approximation also for the family of interpolating Wilson loops consi\-dered in \cite{Polchinski:2011im,Beccaria:2017rbe}.}
\beq
\mbox{log}~W_{\mbox{\scriptsize ren}}^{\mbox{\scriptsize QCD} }~=~0~+~{\cal O}(g^4)~.
\eeq
\section{Conclusions}
In  ${\cal N}=4$ SYM we obtained for the locally supersymmetric as well as for the ordinary Wilson loop
in lowest nontrivial order 
\beq
W_{\mbox{\scriptsize ren}}~=~1 \label{finalW}
\eeq
both at weak and strong coupling.

This result holds also at weak coupling for QCD. Furthermore, it is independent of the angle between the discs spanned by the circles.
Because no logarithmic divergences showed up \footnote{Although present in the supersymmetric case for single spikes at weak coupling, there appears no logarithmic term for the case of two touching spikes. This has been noticed already in  \cite{Dorn:2018als}.}, it is free of any renormalisation group
ambiguity.

Of course the main open question is, whether this result is an accident of 
the lowest orders or whether it extends to all orders. In further work in higher orders
one has to take into account also the mixing with the correlation function for the two
Wilson loops for the single circles. 

Using modifications of {\it AdS}, proposed for holographic QCD, see e.g. \cite{Ammon:2015wua} and references therein, it should be straightforwardly to get the strong coupling result for QCD.

In speculating about physical properties, which could be related to our issue, ones mind is crossed by  
zig-zag 
symmetry \cite{Polyakov:1997tj} and conformal invariance. Zig-zag symmetry means that
a part of a contour which is backtracked contributes  only a factor 1. Classically it is realised
for the ordinary Wilson loops, i.e. gauge parallel transporters, but is violated for the local supersymmetric loop due to the coupling to the scalars, which is not sensitive to the orientation. It is expected
to hold in all orders of perturbation theory for ordinary Wilson loops, and there are arguments, that for the local supersymmetric loops it should be restored in the strong coupling limit \cite{Drukker:1999zq}. 

With this assumption \eqref{finalW} holds as an all order result in QCD for $R_1=R_2$ and $\beta =0$, i.e. the
exact backtracking case. For $R_1>R_2$ there is only local backtracking and the Wilson loop
for the single circles become different due the scale dependence of the renormalised coupling
constant. 

On the other side, in  ${\cal N}=4$ SYM conformal symmetry is unbroken.  The Wilson loops
for single circles are independent of their radius and known as an all order result \cite{Erickson:2000af, Drukker:2000rr}.

A last comment concerns the relation of our result to the symmetry breaking under conformal transformations, which map one point of the contour to infinity. The seminal discussion of this issue in ref. \cite{Drukker:2000rr} applies to cases where the respective point is on a smooth piece of the contour. In our case this point is just the
singular point at the tip of the spikes, i.e. it is not of the type considered in
\cite{Drukker:2000rr} and one should not imperatively expect that their universal anomaly factor
\footnote{It has been derived for Euclidean contours. Variations have been observed also for lightlike polygons \cite{Dorn:2013ita}.} also governs the relation between the touching circles and anti-parallel straight lines. Some details for the comparison
with the case of two anti-parallel lines are collected in appendix C. 
\\[20mm]
{\bf Acknowledgement:}\\[5mm]
I would like to thank the 
Quantum Field and String Theory Group at Humboldt University
for kind hospitality.\\[20mm]
\section*{Appendix A}
This appendix is devoted to the evaluation of the integral
\beq
J(R_1,R_2,\beta,\epsilon)~=~\int_0^{B_{\epsilon}}~\frac{h(R_1,R_2,\beta,\epsilon,x)~dx}{\sqrt{(1-\frac{x^2}{B_{\epsilon}^2})(1+x^4)}  }\label{J}
\eeq
for $\epsilon\rightarrow 0$. $h$ and $B_{\epsilon}$ are defined in \eqref{h}-\eqref{h3} and \eqref{A}, respectively.  

We start with the integral $J_0$  where, compared to $J$,  $h$ is replaced by 1. It can be expressed in terms of the complete elliptic integral $K$ of the first kind via
\bea
J_0(R_1,R_2,\beta,\epsilon)&=&\int_0^{B_{\epsilon}}~\frac{dx}{\sqrt{(1-\frac{x^2}{B_{\epsilon}^2})(1+x^4)}  }~=~\frac{B_{\epsilon}}{\sqrt{1+iB_{\epsilon}}}~K\Big (\frac{2B_{\epsilon}^2}{B_{\epsilon}^2-i}\Big )\label{J0} \\
&=&\mbox{Re}\Big (K\Big(\frac{1}{2}-\frac{i}{2B_{\epsilon}^2}\Big)\Big)~+~\mbox{Im}\Big (K\Big(\frac{1}{2}-\frac{i}{2B_{\epsilon}^2}\Big)\Big)~.
\nonumber
\eea 
$K(y)$ is near $y=1/2$ an analytic function . The deviation from $1/2$ in the second line of \eqref{J0} is proportional to $\epsilon$, see \eqref{A}.
Then expressing $K(1/2)$ in terms of the Gamma function, we get
\beq
J_0(R_1,R_2,\beta,\epsilon)~=~\frac{\big (\Gamma(\frac{1}{4})\big )^2}{4\sqrt{\pi}}~+~{\cal O}(\epsilon)~.\label{J0as}
\eeq
To proceed, we note that the square root factor in the definition of $h$ in \eqref{h} allows an uniform estimate $1+{\cal O}(\epsilon)$. The first factor does not, but is at least bounded in the whole integration interval.  Let us define \footnote{Of course terms containing $\epsilon$ without a factor $x^2$, or $\epsilon^3x^2$ are also irrelevant for our analysis.}
\beq
\hat h_1(R_1,R_2,\beta)~=~\frac{2(R_1^2-R_2^2)+R_{12}^2}{2R_1R_2R_{12}}~,
\eeq
\beq
\hat h_i(R_1,R_2,\beta)~=~h_i(R_1,R_2,\beta)~, ~~~~i=2,3
\eeq
and
\beq
\hat h(R_1,R_2,\beta,\epsilon,x)=\frac{1+\hat h_1~\epsilon x^2+\hat h_2~\epsilon^2x^4}{1+2\hat h_3~\epsilon x^2-\frac{\mbox{\scriptsize sin}^2\beta}{R_{12}^2}\epsilon^2x^4}~,
 \label{hath}
\eeq
as well as 
\beq
\hat J(R_1,R_2,\beta,\epsilon)~=~\int_0^{B_{\epsilon}}~\frac{\hat h(R_1,R_2,\beta,\epsilon,x)~dx}{\sqrt{(1-\frac{x^2}{B_{\epsilon}^2})(1+x^4)}  }~.\label{hatJ}
\eeq
Then we get 
\beq
J(R_1,R_2,\beta,\epsilon)~=~\hat J(R_1,R_2,\beta,\epsilon)~+~{\cal O}(\epsilon)~.\label{JhatJ}
\eeq
Now we split the integration over $x$ in two pieces via
\bea
\hat J(R_1,R_2,\beta,\epsilon)&=&\hat J^{\mbox{\scriptsize lower}}(R_1,R_2,\beta,\epsilon)~+~\hat J^{\mbox{\scriptsize upper}}(R_1,R_2,\beta,\epsilon)\label{hatJsplit}\\
&=&\int_0^{\frac{ b \epsilon^{\alpha}}{\sqrt{\epsilon}}}
~\frac{\hat h(R_1,R_2,\beta,\epsilon,x)~dx}{\sqrt{(1-\frac{x^2}{B_{\epsilon}^2})(1+x^4)}  }
~+~\int_{ \frac{ b \epsilon^{\alpha}}{\sqrt{\epsilon}} }^{B_{\epsilon}}
~\frac{\hat h(R_1,R_2,\beta,\epsilon,x)~dx}{\sqrt{(1-\frac{x^2}{B_{\epsilon}^2})(1+x^4)}}~, \nonumber
\eea
with $b>0$ a fixed number and \footnote{Concerning only $\hat J^{\mbox{\scriptsize lower}}  $,  we could allow $\alpha$ even up to $\frac{1}{2}.$ }
\beq
\frac{1}{4}<\alpha<\frac{3}{8}~.\label{alphafirst}
\eeq
Then the deviation of $\hat h$ from 1 in $\hat J^{\mbox{\scriptsize lower}}  $ is uniformly ${\cal O}(\epsilon^{2\alpha})$, hence
\beq
\hat J^{\mbox{\scriptsize lower}}~=~\int_0^{\frac{ b \epsilon^{\alpha}}{\sqrt{\epsilon}}}
~\frac{dx}{\sqrt{(1-\frac{x^2}{B_{\epsilon}^2})(1+x^4)}  }~+~{\cal O}(\epsilon ^{2\alpha})~.\label{hatJlower}
\eeq
For the estimate of $ \hat J^{\mbox{\scriptsize upper}}  $ we use 
\beq
\frac{1}{\sqrt{1+x^4}}~=~\frac{1}{x^2}\big (1+{\cal O}(\epsilon^{2-4\alpha})\big )~
\eeq
to get with \eqref{hath}, \eqref{hatJsplit}
\bea
\hat J^{\mbox{\scriptsize upper}}&=&\int^{B_{\epsilon}}_{\frac{ b \epsilon^{\alpha}}{\sqrt{\epsilon}}}
~\frac{\hat h(R_1,R_2,\beta,\epsilon,x)~dx}{x^2~\sqrt{1-\frac{x^2}{B_{\epsilon}^2} }}~+~{\cal O}(\epsilon ^{2-4\alpha})\nonumber\\
&=&\int^{B_{\epsilon}}_{\frac{ b \epsilon^{\alpha}}{\sqrt{\epsilon}}}
~\frac{dx}{x^2~\sqrt{1-\frac{x^2}{B_{\epsilon}^2} }}~+~V_1~+V_2~+~{\cal O}(\epsilon ^{2-4\alpha})~,\label{hatJupper2}
\eea
where
\bea
V_1(R_1,R_2,\beta,\epsilon)&=&(\hat h_1-2\hat h_3)~\epsilon~ \int^{B_{\epsilon}}_{\frac{ b \epsilon^{\alpha}}{\sqrt{\epsilon}}}
~\frac{dx}{\sqrt{1-\frac{x^2}{B_{\epsilon}^2} }~~\big (1+2h_3 \epsilon x^2-\frac{\mbox{\scriptsize sin}^2\beta}{R_{12}^2}\epsilon^2x^4\big )}~,\\
V_2(R_1,R_2,\beta,\epsilon)&=&\big (\hat h_2+\frac{\sin^2\beta}{R^2_{12}}\big )~\epsilon^2 ~\int^{B_{\epsilon}}_{\frac{ b \epsilon^{\alpha}}{\sqrt{\epsilon}}}
~\frac{x^2~dx}{\sqrt{1-\frac{x^2}{B_{\epsilon}^2} }~~\big (1+2h_3\epsilon x^2-\frac{\mbox{\scriptsize sin}^2\beta}{R_{12}^2}\epsilon^2x^4\big )}~.\nonumber
\eea
Adding \eqref{hatJlower} and \eqref{hatJupper2} we can reinstall the factor $1/\sqrt{1+x^4}$ instead of $1/x^2$ in the first term on the r.h.s. of \eqref{hatJupper2} and arrive with \eqref{hatJsplit}, \eqref{J0} and \eqref{JhatJ} at
\beq
J(R_1,R_2,\beta,\epsilon)=J_0~+~\sum_{n=1}^2 V_n(R_1,R_2,\beta,\epsilon)+{\cal O}(\epsilon)+{\cal O}(\epsilon^{2-4\alpha})+{\cal O}(\epsilon ^{2\alpha})~.\label{JJ0U}
\eeq
The integrals in both $V_1$ and $V_2$ can be expressed in terms of inverse trigonometric functions, and after some algebra we get
\bea
V_1(R_1,R_2,\epsilon)&=&-\frac{\pi \sqrt{\epsilon}}{4}~\frac{\sqrt{2R_1R_2R_{12}}}{\vert R_{12}^2-R_1^2\vert~ }~\cdot\left (\frac{R_{12}\cos\beta}{R_1} ~\Theta\big (\cos\beta-\frac{R_2}{2R_1}\big )\right .\\[2mm]
&&~~~~~~~~~~~~~~~~~~~~~~~~~~~~~\left .+\big (\frac{R_2}{R_1}-\cos\beta\big )~\Theta\big (\frac{R_2}{2R_1}-\cos\beta\big)\right )~+~{\cal O}(\epsilon^{\frac{1}{2}+\alpha})~,\nonumber\\[2mm]
V_2(R_1,R_2,\epsilon)&=&\frac{\pi \sqrt{\epsilon}}{4}~\frac{\sqrt{2R_1R_2R_{12}}}{\vert R_{12}^2-R_1^2\vert }~\\[2mm]
&&~~~\cdot\left (\frac{R_{12}\cos\beta}{R_1}~\Theta\big (\cos\beta-\frac{R_2}{2R_1}\big)+\cos\beta~\Theta\big (\frac{R_2}{2R_1}-\cos\beta\big )\right )~+~{\cal O}(\epsilon^{\frac{1}{2}+\alpha})\nonumber ~.\eea
This implies
\beq 
\sum_{n=1}^2V_n~=~-\frac{\pi \sqrt{\epsilon}}{4}~\sqrt{\frac{2R_{12}}{R_1R_2}}~\Theta\big (\frac{R_2}{2R_1}-\cos\beta\big)~+~{\cal O}(\epsilon^{\frac{1}{2}+\alpha})~.
\eeq
Inserting this in \eqref{JJ0U} and using \eqref{J0as} as well as \eqref{alphafirst} we arrive at
\footnote{By $o(\sqrt{\epsilon})$ we denote terms vanishing {\it faster} than $\sqrt{\epsilon}$.}
\beq
J(R_1,R_2,\epsilon)~=~\frac{(\Gamma(\frac{1}{4}))^2}{4\sqrt{\pi}}~-~\frac{\pi \sqrt{\epsilon}}{4}~\sqrt{\frac{2R_{12}}{R_1R_2}}~\Theta\big (\frac{R_2}{2R_1}-\cos\beta\big) ~+~o(\sqrt{\epsilon})~.\label{appAfinal}
\eeq
\section*{Appendix B}
We need the $\epsilon\rightarrow 0$ behaviour of the integrals \eqref{Aeps}.
The corresponding analysis follows closely the lines of \cite{Dorn:2018als}, see also footnote \footnote{An error in the journal version has been indicated in the erratum enclosed in the citation. The related correction of the relevant appendix can be found in the updated arXiv version.}.
Nevertheless there are two reasons to present it here in detail. 
At first with the angle
$\beta$ an additional new parameter is present, and secondly we now have to be
more careful, since we need also the finite term, which was not of interest in \cite{Dorn:2018als}.

Adding  under the square root in the nominator of the integrand a zero in the form of the l.h.s. of \eqref{region} we get
\beq
A_{\epsilon}^{\pm}~=~\frac{2~r_0^2}{\sqrt{\epsilon}}~\int_{r_{\epsilon}^{\pm}}^{r_0}\frac{\sqrt{r-r_{\epsilon}^{\pm}}\sqrt{1-\epsilon~ f_{\pm}(r,,r_{\epsilon}^{\pm})}}{r^2~\sqrt{r_0^4-r^4}}~dr~,\label{A-app-B}
\eeq
with
\beq
f_{\pm}(r,r_{\epsilon}^{\pm})~=~ \frac{(M\pm \sigma(r))^2-(M\pm \sigma (r_{\epsilon}^{\pm}))^2}{r-r_{\epsilon}^{\pm}}~+~ (r+r_{\epsilon}^{\pm}) ~.\label{f}
\eeq
This has the same form as the corresponding equation for $\beta=0$ in \cite{Dorn:2018als}. The $\beta$-dependence enters here only via that of $M$, see \eqref{M-def}. There is more $\beta$-dependence in the equation for $r_{\epsilon}^{\pm}$, \eqref{region}. But for the expansion at small $\epsilon$ we can use  
\beq
(M\pm L/2)^2~+~\frac{\sin^2\beta}{4R_{12}^2}~=~\frac{1}{4R_{2}^2}~,~~\mbox{or}~~\frac{1}{4R_{1}^2}~,
\eeq
which follows from \eqref{R12},\eqref{L} and \eqref{M-def}. Then with  \eqref{sigma-r} and \eqref{r0L} we get
\bea
r_{\epsilon}^{+}&=&\frac{\epsilon}{4R_2^2}~+~\frac{\epsilon^3}{16R_2^4}~+~{\cal O}(\epsilon ^4)~,\nonumber\\
r_{\epsilon}^{-}&=&\frac{\epsilon}{4R_1^2}~+~\frac{\epsilon^3}{16R_1^4}~+~{\cal O}(\epsilon ^4)~.
\label{repsas}
\eea
Note that these expansions up to terms $\propto \epsilon^3$ do not depend on $\beta$.

Now we split $A_{\epsilon}^{\pm}$ in two pieces
\beq
A_{\epsilon}^{\pm}~=~A_{\epsilon,\mbox{\scriptsize lead}}^{\pm}~+~A_{\epsilon,\mbox{\scriptsize rem}}^{\pm}~,\label{Asplit}
\eeq
with
\beq
A_{\epsilon,\mbox{\scriptsize lead}}^{\pm}~=\frac{2}{\sqrt{\epsilon~r_0}}~\int_{r_\epsilon^{\pm}/r_0}^{1}\frac{\sqrt{x-r_{\epsilon}^{\pm}/r_0}}{x^2\sqrt{1-x^4}}~dx\label{lead}
\eeq
and
\beq
A_{\epsilon,\mbox{\scriptsize rem}}^{\pm}~=~\frac{2}{\sqrt{\epsilon~r_0}}~\int_{r_\epsilon^{\pm}/r_0}^{1}\frac{\sqrt{x-r_{\epsilon}^{\pm}/r_0}}{x^2\sqrt{1-x^4}}~\Big (\sqrt{1-\epsilon~f_{\pm}(xr_0,r_{\epsilon}^{\pm})}~-~1\Big )~dx~.\label{rem}
\eeq
For the estimate of \eqref{lead} we use
\bea
\int_{\delta}^1\frac{\sqrt{x-\delta}}{x^2\sqrt{1-x^4}}dx&=&\frac{\pi}{2\sqrt{\delta}}~-~\frac{2\sqrt{\pi}~\Gamma(\frac{7}{8})}{\Gamma(\frac{3}{8})}~_5F_4\Big (-\frac{1} {8},\frac{1}{8},\frac{3}{8},\frac{5}{8},\frac{5}{8}; \frac{1}{4},\frac{1}{2},\frac{3}{4},\frac{9}{8};\delta^4\Big )\nonumber\\[2mm]
&&~~~~~~~~+\delta~b_1(\delta^4)+\delta^2~b_2(\delta^4)+\delta^3~b_3(\delta^4)~,\label{5F4}
\eea
where $b_1,b_2,b_3$ are given by different hypergeometric $_5F_4$'s of argument $\delta^4$ times some numerical factors. With \eqref{repsas} this implies 
($j(+)=2,~j(-)=1$)
\beq
A_{\epsilon,\mbox{\scriptsize lead}}^{\pm}~=~\frac{2\pi R_{j(\pm)}}{\epsilon}~-~\frac{4\sqrt{\pi}~\Gamma(\frac{7}{8})}{\Gamma(\frac{3}{8})}~\frac{1}{\sqrt{\epsilon~r_0}}~+~{\cal O}(\sqrt{\epsilon})~.\label{Alead}
\eeq
Concerning the estimate of $A_{\epsilon,\mbox{\scriptsize rem}}^{\pm}$ we noted in \cite{Dorn:2018als}, that $f_{\pm}(r,r_{\epsilon}^{\pm})$ is bounded for $\epsilon\rightarrow 0$ uniformly with respect to $r$. This allowed to conclude
\beq
A_{\epsilon,\mbox{\scriptsize rem}}^{\pm}~=~{\cal O}(\epsilon)\cdot A_{\epsilon,\mbox{\scriptsize lead}}^{\pm}~=~{\cal O}(\epsilon^0)~.
\eeq
But we can be more efficiently. Expanding the square root in \eqref{rem} and using again the uniform boundedness of $f_{\pm}(r,r_{\epsilon}^{\pm})$ we get
\beq
A_{\epsilon,\mbox{\scriptsize rem}}^{\pm}~=~-\frac{\epsilon}{\sqrt{\epsilon~r_0}}~\int_{r_\epsilon^{\pm}/r_0}^{1}\frac{\sqrt{x-r_{\epsilon}^{\pm}/r_0}}{x^2\sqrt{1-x^4}}~f_{\pm}(xr_0,r_{\epsilon}^{\pm})dx~+~{\cal O}(\epsilon^2)A_{\epsilon,\mbox{\scriptsize lead}}^{\pm}~.
\eeq
The integral without the factor  $f_{\pm}(r,r_{\epsilon}^{\pm})$ would diverge for $\epsilon\rightarrow 0$ according to \eqref{5F4}. But due to the behaviour
\beq
f_{\pm}(r,r_{\epsilon}^{\pm})~=~{\cal O}(\epsilon)~+~\big (1+{\cal O}(\epsilon)\big )(r-r_{\epsilon}^{\pm})~+~{\cal O}((r-r_{\epsilon}^{\pm})^2)
\eeq
for small $\epsilon$, and near the lower boundary of the integral, it remains finite. This means
\beq
A_{\epsilon,\mbox{\scriptsize rem}}^{\pm}~=~{\cal O}(\sqrt{\epsilon})
\eeq
and together with \eqref{Alead},\eqref{Asplit},\eqref{A-Apm}
\beq
A_{\epsilon}~=~\frac{2\pi (R_1+R_2)}{\epsilon}~-~\frac{8~(2\pi)^{\frac{5}{4}}~\Gamma(\frac{7}{8})}{\Gamma(\frac{3}{8})\Gamma(\frac{1}{4})}~\frac{1}{\sqrt{\epsilon~\vert \vec k_1-\vec k_2\vert }}~+~{\cal O}(\sqrt{\epsilon})~.\label{appBfinal}
\eeq
Here use has been made also of the relations between $r_0,~L$ and the curvature difference $\vert \vec k_1-\vec k_2\vert$, i.e. \eqref{r0L},\eqref{L} and \eqref{R12-curv}.
\section*{Appendix C}
Here we collect some details for the comparison of the two touching circles with two antiparallel straight lines. 
Performing the trivial integrations for lines at distance $L$ one
gets 
\bea
\mbox{log} W_{\mbox{\scriptsize parallel}}&=&\frac{g^2C_F}{2\pi^2}\Big (\frac{2l}{L}~\mbox{arctan}\frac{l}{L}~-~\mbox{log}(1+\frac{l^2}{L^2})\Big )~+~{\cal O}(g^4)~,
\nonumber \\
&=&\frac{g^2C_F}{2\pi^2}\Big (\frac{\pi l}{L}~-~2~\mbox{log}\frac{l}{L}~-2~+~{\cal O}(\frac{L^2}{l^2})\Big )~+~{\cal O}(g^4)~.\label{parallel}
\eea
To control the infrared problem, the integration has been restricted to straight
lines of length $l$, with the goal $l\rightarrow\infty$. Contrary to the treatment of
ultraviolet divergences, there is no recipe to give for infinitely
extended contours  the Wilson loop a finite meaning per se. Nevertheless it is the source for a
meaningful physical quantity, the static quark-antiquark potential, via 
$V(L)~=~$lim$_{l\rightarrow\infty} ~W_{\mbox{\scriptsize parallel}}/l $.
Thus this potential is just given by the factor of the linear infrared divergence.

The Wilson loop for the touching circles is  from \eqref{logW},\eqref{Ijfinal}, \eqref{R12-curv} and \eqref{I12final}
\beq
\mbox{log}W~=~\frac{g^2C_F}{4\pi^2}\Big (\sqrt{2\pi}\Big (\Gamma(\frac{1}{4})\Big )^2\frac{1}{\sqrt{\epsilon \vert \vec k_1-\vec k_2\vert}}~+~o(\epsilon ^0)\Big )~+~{\cal O}(g^4)~.
\eeq
Our ultraviolet regularisation parameter $\epsilon$, as used in chapter 2, mimics
a universal cutoff in the distance between the two endpoints of the propagator.

The special situation near the touching point of the two circles could be regularised
also by restricting the integrations to the image under inversion of the two
straight lines of finite length \nolinebreak  l. Then the minimum of the allowed propagator
distances would be
\beq
\mbox{min}\vert  \vec x_1-\vec x_2\vert~=~\frac{2 R_{12}}{\sqrt{(1+l^2R_1^2)(1+l^2R_2^2)}}~=~\frac{2~\vert \vec k_1-\vec k_2\vert}{l^2}~+~{\cal O}\big (\frac{1}{l^4}\big )~.
\eeq
Identifying this minimum with $\epsilon$ one finds, starting from \eqref{parallel},
the spiky ultraviolet $1/\sqrt{\epsilon}$ divergence as an image of the linear infrared divergence. But invariance under inversion is broken, resulting in different numerical
coefficients. Furthermore, there are different finite terms and no logarithmic divergence for the circles.

Of course, this interplay between the IR for straight lines and the UV for the circles holds also for strong coupling. It is illustrated in an eye-catching manner
in figure 3 of \cite{Dorn:2018als}. But due to symmetry breaking, also here the
coefficients require independent calculations.

\end{document}